\documentclass[12pt]{article}
\pdfoutput=1

\usepackage[totalheight = 23cm, totalwidth = 17cm]{geometry}
\usepackage{amssymb,amsmath,amsfonts,amsbsy,graphicx}
\usepackage{color,comment,url}
\usepackage{hyperref}

\usepackage{subfigure}
\usepackage{ulem}
\usepackage{cite}

\newcommand{\mpl}{m_{\rm Pl}}

\definecolor{purple}{rgb}{0.5 ,0, 0.7}
\definecolor{bluegreen}{rgb}{0, 0.45, 0.35}
\hypersetup{colorlinks=true, linkcolor=bluegreen, citecolor=purple, urlcolor=blue}

\allowdisplaybreaks

\begin{document}

\begin{titlepage}

\begin{center}

\leavevmode \\
\vspace{ 0cm}

\hfill APCTP Pre2016-022 \\
\hfill KIAS-P16084

\vskip 2 cm

{\LARGE \bf Curvaton as dark matter \\ with secondary inflation}

\vglue .6in

{\large
 Jinn-Ouk Gong$^{a,b}$, Naoya Kitajima$^{a}$ and Takahiro Terada$^{c}$
}

\vglue.3in

\textit{
   ${}^{a}$ Asia Pacific Center for Theoretical Physics, Pohang 37673, Korea
   \\
   ${}^{b}$ Department of Physics, POSTECH, Pohang 37673, Korea
   \\
   ${}^{c}$ Korea Institute for Advanced Study, Seoul 02455, Korea
}

\end{center}

\vglue 0.6in

\begin{abstract}

We consider a novel cosmological scenario in which a curvaton is long-lived and plays the role of cold dark matter (CDM) in the presence of a short, secondary inflation. Non-trivial evolution of the large scale cosmological perturbation in the curvaton scenario can affect the duration of the short term inflation, resulting in the inhomogeneous end of inflation. Non-linear parameters of the curvature perturbation are predicted to be $f_{\rm NL} \approx 5/4$ and $g_{\rm NL} \approx 0$. The curvaton abundance can be well diluted by the short-term inflation and accordingly, it does not have to decay into the Standard Model particles. Then the curvaton can account for the present CDM with the isocurvature perturbation being sufficiently suppressed because both the adiabatic and CDM isocurvature perturbations have the same origin. As an explicit example, we consider the thermal inflation scenario and a string axion as a candidate for this curvaton-dark matter. We further discuss possibilities to identify the curvaton-dark matter with the QCD axion.

\end{abstract}

\end{titlepage}

\section{Introduction}

The existence of non-vanishing Sachs-Wolfe plateau~\cite{Sachs:1967er} in the power spectrum of the temperature anisotropies of the cosmic microwave background (CMB)~\cite{Bennett:1996ce} indicates pre-existing perturbations on super-horizon scales before the onset of the hot big bang evolution of the universe. This calls for, unless finely tuned by coincidence, a dynamical explanation for these perturbations. The most popular mechanism for the causal generation of the primordial perturbations is  cosmic inflation~\cite{inflation}. In the simplest realization, the quantum fluctuations of the light inflaton field are responsible for the primordial curvature perturbation that becomes the seed of subsequent inhomogeneities including the CMB temperature fluctuations~\cite{books}. The power spectrum of the curvature perturbation is predicted to be nearly scale-invariant and almost Gaussian, which is consistent with observations~\cite{Ade:2015lrj}.

The inflaton, however, needs not be the only matter content in the universe before the standard hot big bang evolution begins. Yet elusive theories of high energy physics, relevant for the very early universe, allow the existence of many degrees of freedom, and they can play certain roles after inflation. A representative example is the so-called curvaton~\cite{curvaton}, a light scalar field with quantum fluctuations of order of the inflationary Hubble parameter: the curvaton can contribute to the primordial perturbation through coherent oscillations after inflation in the thermal background produced by the inflaton. Furthermore, those other degrees of freedom after inflation can also give rise to diverse possibilities of thermal history, leading to solutions for the potential problems such as the moduli overproduction~\cite{moduli}. A short, secondary inflation, with thermal inflation~\cite{Lyth:1995hj} being a good example, driven by another field can dilute otherwise dangerous relics.

Scalar degrees of freedom are ubiquitous in high energy particle physics beyond the Standard Model. For example, the QCD axion~\cite{axion} is a pseudo Nambu-Goldstone boson associated with the spontaneous breaking of the Peccei-Quinn (PQ) symmetry~\cite{Peccei:1977hh} (for recent reviews, see e.g. Ref.~\cite{axion-review}). String theory also predicts plenty of light scalar fields. Some of them are called string axions and they often play important roles in cosmology as being known as the string axiverse~\cite{axiverse}. These axions can be plausible candidates for the present cold dark matter (CDM) because they are typically long-lived and have only extremely weak interactions with the Standard Model particles~\cite{misalignment}.

The presence of axions can significantly affect the history of the early universe. For example, axions acquire quantum fluctuations during inflation, which results in the CDM isocurvature mode in the CMB temperature anisotropy if axions contribute to the present CDM component. Current observations put a tight upper bound~\cite{Ade:2015lrj} and the axion dark matter is especially incompatible with high-scale inflation in the case where the PQ symmetry is broken before or during inflation. In the case of relatively heavy axion which decays in some early stages, the isocurvature perturbation is converted to the adiabatic perturbation. In such a case, axion takes a role of the curvaton~\cite{Dimopoulos:2003az}.

In this article, we consider a novel curvaton scenario in which the curvaton is long-lived and constitutes the present CDM, in stark contrast to the standard curvaton scenario in which the curvaton must decay to realize the standard radiation-dominated universe.  Usually, the curvaton and CDM cannot be a single field. However, we point out that it is possible by invoking a late-time secondary, short-term inflation like thermal inflation. Due to the mechanism we explain in detail below, the curvaton fluctuation on the initial flat hypersurface affects the duration of the secondary inflation and eventually it is copied to the fluctuation of the secondary inflaton component. In particular, our scenario does not require any interactions between the curvaton and the Standard Model sector except gravity. 
Our scenario can be straightforwardly applicable to the axion case because it has suitable features for both the curvaton and CDM as mentioned above. The isocurvature perturbation can be suppressed regardless of the (primordial) inflation scale since both of the perturbations of radiation and CDM (curvaton) have the same origin. It can be an alternative idea to avoid the isocurvature perturbation problem for the axion CDM. Thus, we may summarize the role of our axion particle schematically as
\begin{equation*}
\text{curvaton} = \text{dark matter} = \text{axion-like particle}
\end{equation*}
in this economical scenario. We explain our main idea in more detail in Section~\ref{sec:2ndinf-curvaton}. As a well-motivated and explicit example, we focus on string axions as the curvaton-dark matter in Section~\ref{sec:ALP}. Section~\ref{sec:conclusion} is our conclusions along with brief discussions.

\section{Secondary inflation and curvaton}
\label{sec:2ndinf-curvaton}

In this section, we sketch the main idea. For definiteness, as the secondary phase of inflation, we consider thermal inflation proposed in \cite{Lyth:1995hj} (for earlier related studies, see \cite{Yamamoto:1985rd}) and extended for moduli problem in more realistic supersymmetric models \cite{Asaka:1997rv}. But as we will discuss later, any specific model of inflation respecting our assumptions would work.

\subsection{Background dynamics}
\label{ssec:bkg}

Thermal inflation is driven by an additional scalar field $\phi$ called flaton, whose potential can be written as
\begin{equation}
V(\phi) = V_0 + \frac{1}{2} \left( gT^2 - m_\phi^2 \right)\phi^2 + \lambda\frac{\phi^n}{M^{n-4}} \, .
\end{equation}
Here, $V_0$ is the vacuum energy, $T$ is the cosmic temperature, $g$ is the coupling between the flaton and the thermal bath, $m_\phi$ is the flaton bare mass, $M$ is some energy scale related to the vacuum expectation value of the flaton, and $\lambda$ is a self coupling constant. We set $g=\lambda = 1$ for simplicity. At high temperature, the flaton is stabilized at the origin due to its thermal mass. When the energy density of the radiation-curvaton fluid becomes smaller than $V_0$, i.e. $\rho_r + \rho_\sigma = V_0$, thermal inflation takes place. It lasts until the temperature decreases to the flaton bare mass and ends suddenly at that time because of tachyonic instability. The $e$-folding number during thermal inflation is thus estimated by
\begin{equation}
N_{\rm TI} = \log \left( \frac{T_{\rm start}}{m_\phi} \right) 
= \frac{1}{4} \log \left[ \left( \frac{60}{\pi^2 g_{* \text{start}}} \right) 
\frac{\Omega_r(t_{\rm start})V_0}{m_\phi^4} \right] \, ,
\end{equation}
where the subscript ``start'' means that the value is evaluated at the onset of thermal inflation\footnote{Here and in what follows, we neglect the change of the relativistic degrees of freedom from the beginning of the curvaton oscillation to the end of thermal inflation in favor of notational simplicity, $g_{*\text{osc}} = g_{*\text{start}} = g_{*\text{end}} \equiv g_*$. It is easy to restore the precise dependence on them, but it does not affect the quantitative results significantly.}. Typically, $N_{\rm TI} \sim 5 - 10$. After the end of thermal inflation, the flaton starts coherent oscillations around its true vacuum and behaves like pressureless matter. It eventually decays and produces a huge amount of entropy so that any pre-existing components are diluted.

The curvaton $\sigma$ is also significantly diluted if its oscillation begins before the onset of thermal inflation, i.e. $m_\sigma > H_{\rm TI}$.   It implies that the curvaton can be long-lived and thus can serve as a CDM component in the present universe. To estimate the current density parameter of the curvaton, it is convenient to introduce a dilution factor $\Delta$ defined by the ratio of the entropy densities just before and just after the flaton decay \cite{Lyth:1995hj,Asaka:1997rv},
\begin{align}
\label{eq:dilution}
\Delta \equiv & \frac{s_{\rm after}}{s_{\rm before}}
= \frac{m_\phi e^{4N_{\rm TI}}}{2 \Omega_r (t_{\rm start}) T_R} \, ,
\end{align}
where $T_R$ is the second reheating temperature defined by the temperature just after the flaton decay. With the curvaton oscillating well before thermal inflation, the ratio of the present curvaton density to the entropy density is calculated as\footnote{We assume that the potential minimum of the curvaton does not change before and after  thermal inflation so that there is no secondary coherent oscillation after thermal inflation. This is true in the string axion case~\cite{Kawasaki:2014una}.}
\begin{equation}
\frac{\rho_\sigma}{s} =  \frac{3 T_{\rm osc} \Omega_{\sigma,{\rm osc}}}{4 \Delta} = 
\frac{1}{8\Delta} \left( \frac{\pi^2g_*}{90} \right)^{-1/4} \sqrt{m_\sigma\mpl}
\left( \frac{\sigma_i}{\mpl} \right)^2 \, ,
\end{equation}
where $T_{\rm osc}$ and $\Omega_{\sigma,{\rm osc}}$ are respectively the temperature and the density parameter of the curvaton at the onset of the curvaton oscillation given by
\begin{equation}
T_{\rm osc} =  \left( \frac{\pi^2g_*}{90} \right)^{-1/4} \sqrt{m_\sigma\mpl}
\quad \text{and} \quad
\Omega_{\sigma,{\rm osc}} = \frac{1}{6} \bigg(\frac{\sigma_i}{\mpl} \bigg)^2 \, , 
\end{equation}
and $\sigma_i$ is the initial amplitude of the curvaton. Therefore we obtain the present value of the curvaton density parameter as
\begin{equation}
\label{eq:cdm_abundance}
\Omega_{\sigma} h^2 \approx 2 \times  \frac{10^{18}}{\Delta} \left( \frac{m_\sigma}{1~{\rm TeV}} \right)^{1/2} \left( \frac{\sigma_i}{\mpl} \right)^2 \, .
\end{equation}
Thus we see that it is easy to realize $\Omega_\sigma \ll 1$ mainly due to the large dilution factor, satisfying the current upper bound, $\Omega_\sigma h^2 < \Omega_{\rm CDM}h^2 = 0.11$. This strong dilution of any pre-existing, potentially dangerous component is the original motivation of thermal inflation, or more generally any short secondary inflation.

The fraction of the radiation component at the beginning of thermal inflation can be calculated as
\begin{equation}
\Omega_r(t_{\rm start}) = 
\frac{1}{2} \left( \Omega_{\sigma,\text{osc}} \frac{T_{\rm osc}}{m_\phi} e^{-N_{\rm TI}} + 1 \right)^{-1} \, .
\end{equation}
Using eq.~\eqref{eq:dilution} with $\Omega_r(t_{\rm start})$ above, we can approximate eq.~\eqref{eq:cdm_abundance} as
\begin{equation}
\label{eq:cdm_abundance_2}
\Omega_{\sigma} h^2 \approx 
\begin{cases} 
2 \times 10^{15} e^{-4N_{\rm TI}} \left( \cfrac{T_R}{1~{\rm GeV}} \right) \left( \cfrac{1~{\rm TeV}}{m_\phi} \right) 
\left( \cfrac{m_\sigma}{1~{\rm TeV}} \right)^{1/2} \left( \cfrac{\sigma_i}{\mpl} \right)^2 
&   \text{ for } \quad \Omega_r(t_{\text{start}}) \gg \Omega_{\sigma}(t_{\text{start}})
\\[4mm]
5 \times 10^8 e^{-3N_{\rm TI}} \left( \cfrac{g_*}{106.75} \right)^{1/4} \left( \cfrac{T_R}{1~{\rm GeV}} \right) 
& \text{ for } \quad \Omega_r(t_{\text{start}}) \ll \Omega_{\sigma}(t_{\text{start}})
\end{cases}
\, .
\end{equation}
Note that for the second case, the final abundance does not depend on $m_\phi$, $m_\sigma$ and $\sigma_i$ but depends only on $N_{\rm TI}$ and $T_R$.

\subsection{Evolution of perturbations}

Now we consider how the initial curvaton fluctuation contributes to the final adiabatic curvature perturbation. In our analysis, we make the following two assumptions:
\begin{description}
\item[(i)] The intrinsic flaton fluctuation is negligible and thermal inflation is completely homogeneous, at least on large scales.
\item[(ii)] The transition between thermal inflation and the subsequent flaton oscillation occurs suddenly.
\end{description}
The first assumption is in fact a very good approximation, as the pre-existing fluctuations of the flaton are lost once thermal inflation begins when the flaton is trapped in the false vacuum due to thermal mass. Furthermore, both the vacuum fluctuations of the flaton and the temperature fluctuations during thermal inflation are suppressed on large scales, with their power spectra being steeply blue~\cite{thermal-inf_pert}. Thus during thermal inflation, we regard the flaton as homogeneous. The second assumption means that even if the flaton is homogeneous, the end of thermal inflation is set by the cosmic fluid other than the flaton, which does have perturbations, so that the curvature perturbation is suddenly transferred to the flaton sector at the end of thermal inflation hypersurface. Then after the decay of the flaton, the flaton fluctuation is eventually converted to that in radiation, which gives adiabatic perturbation.

\begin{figure}[ht!]
 \centering
 \includegraphics[width=0.7\textwidth]{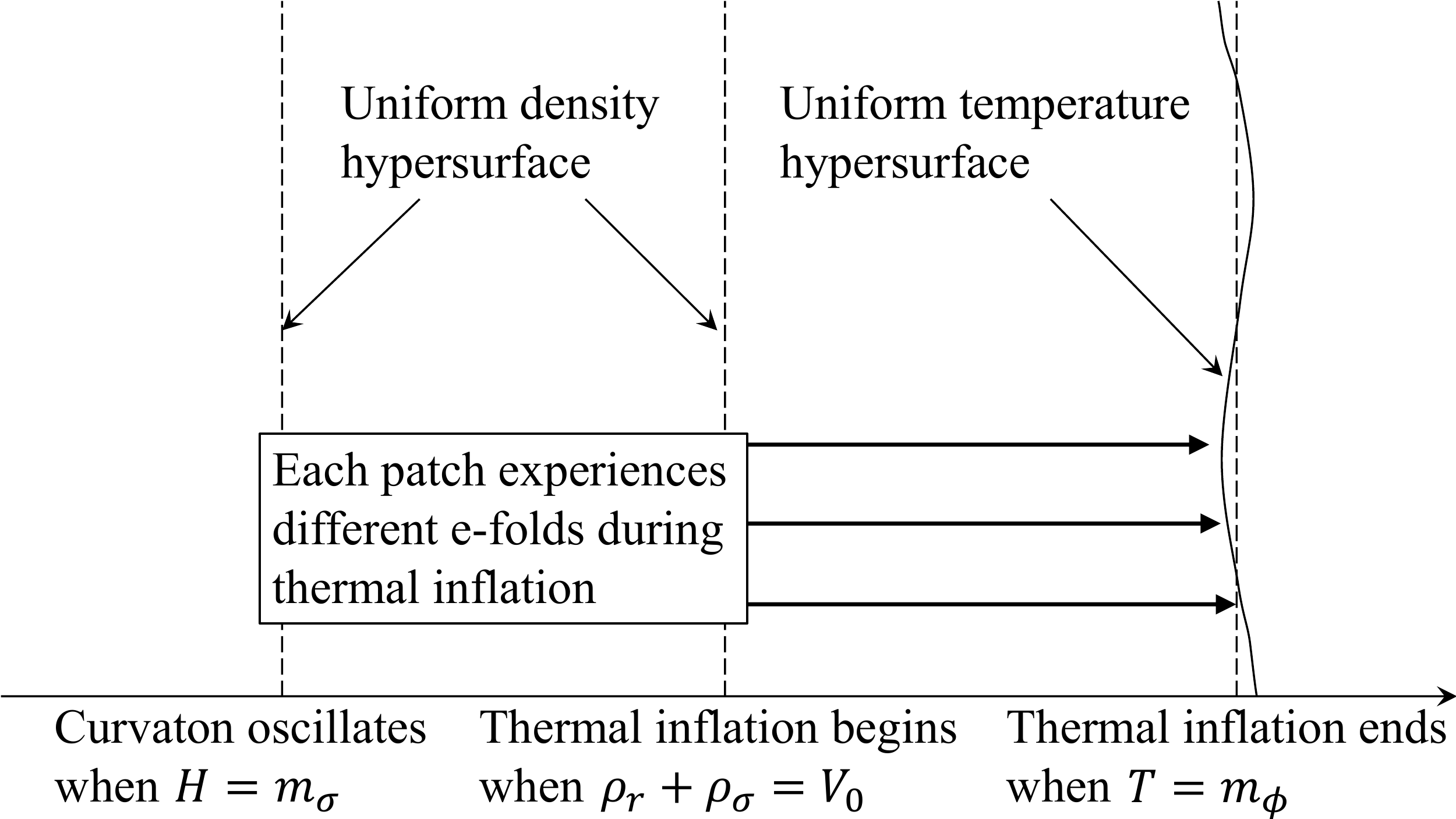}
 \caption{A schematic plot of the evolution of the universe including thermal inflation. While the oscillation of the curvaton and the onset of thermal inflation occur on the uniform density hypersurfaces, the end of thermal inflation is set by temperature and thus occurs on the uniform temperature hypersurface. This leads to different $e$-foldings during thermal inflation in each patch of the universe.}
 \label{fig:expansion}
\end{figure}

Following the evolution of the large-scale perturbations on the uniform density hypersurface, the curvaton oscillation and the beginning of thermal inflation occurs uniformly. However, since the end of thermal inflation depends only on the temperature, this hypersurface does not coincide with the uniform density hypersurface. In other words, the $e$-folding number during thermal inflation is different in each patch of the universe due to the inhomogeneous end of inflation~\cite{Lyth:2005qk}. That is why the final adiabatic perturbation can be generated by the curvaton which is completely decoupled from the visible sector. This is schematically shown in Figure~\ref{fig:expansion}.

Let us begin with the non-linear curvature perturbation associated with each cosmic fluid labeled by the subscript $X$~\cite{nl-zeta}:
\begin{equation} 
\label{zeta_X}
\zeta_X = \delta{N} + \frac{1}{3(1+w_X)} \log \left( \frac{\rho_X}{\bar\rho_X} \right) \, ,
\end{equation}
where $w_X$ is the equation-of-state parameter, and $\rho_X$ ($\bar\rho_X$) denotes the (background) energy density of $X$. Then in a flat Friedmann-Lema\^{i}tre-Robertson-Walker universe, the energy conservation in each patch of the universe tells us the non-trivial super-horizon evolution for the curvature perturbation on arbitrary hypersurfaces,
\begin{equation}
\label{eq:E-cons}
\sum_X \Omega_X e^{3(1+w_X)(\zeta_X-\delta N)} = e^{3(1+w_{\rm tot})(\zeta-\delta N)} \, ,
\end{equation}
where $\Omega_X = \bar\rho_X/\bar\rho_{\rm tot}$ is the background density parameter for each component and $\zeta$ is the curvature perturbation on the uniform total density hypersurface.

We now apply eq.~\eqref{eq:E-cons} to the case of our interest. Because thermal inflation ends suddenly at $T = m_\phi$, the hypersurface of the uniform end of thermal inflation coincides with the uniform temperature hypersurface, that is $\delta N = \zeta_r$. On this hypersurface, we can obtain the curvature perturbation just before the end of thermal inflation via
\begin{equation}
\label{eq:beforeTI}
\Omega_\phi + \Omega_\sigma e^{3(\zeta_\sigma-\zeta_r)} + \Omega_r = e^{3(1+w_{\rm tot})(\zeta-\zeta_r)} \, ,
\end{equation}
where the total equation of state is given by
\begin{equation}
3(1+w_{\rm tot}) = 3\Omega_\sigma + 4\Omega_r \, ,
\end{equation}
with the use of $w_\phi = -1$. Note that the flaton component is completely homogeneous by our assumption (i). Expanding each exponential up to the linear order, the total curvature perturbation is solved as
\begin{equation}
\zeta = \zeta_r + \frac{r_\sigma(t_{\rm end})}{3}S_\sigma \, ,
\end{equation}
where $S_\sigma \equiv 3(\zeta_\sigma-\zeta_r)$ is the isocurvature perturbation of the curvaton relative to radiation, and
\begin{equation}
\label{eq:r_sigma}
r_\sigma(t_{\rm end}) \equiv \frac{3\Omega_\sigma}{4\Omega_r + 3\Omega_\sigma}\bigg|_{t = t_{\rm end}} \, ,
\end{equation}
which is evaluated at the end of thermal inflation.

From the assumption (ii), the equation of state for the flaton changes suddenly from $w_\phi = -1$ to $w_\phi = 0$ at the end of thermal inflation. Then, eq.~\eqref{eq:E-cons} right after the end of thermal inflation on $T=m_\sigma$ slicing gives
\begin{equation}
\label{eq:afterTI}
\Omega_\phi e^{3(\zeta_\phi-\zeta_r)} + \Omega_\sigma e^{3(\zeta_\sigma-\zeta_r)} + \Omega_r = e^{3(1+w_{\rm tot})(\zeta-\zeta_r)} \, ,
\end{equation}
where the equation of state parameter is now given by
\begin{equation}
3(1+w_{\rm tot}) = 3\Omega_\phi + 3\Omega_\sigma + 4\Omega_r \approx 3\Omega_\phi \, ,
\end{equation}
and by expanding exponentials up to the linear order in perturbations we obtain
\begin{equation}
\zeta = \frac{3\Omega_\phi \zeta_\phi + 3\Omega_\sigma \zeta_\sigma + 4\Omega_r \zeta_r}{3\Omega_\phi + \Omega_\sigma + 4\Omega_r} \approx \zeta_\phi \, .
\end{equation}
Since the value of $\zeta$ should be continuous, by matching the values before and after the transition we find 
\begin{equation}
\label{eq:zeta_flaton}
\zeta_\phi \approx  \zeta = \zeta_r + \frac{r_\sigma(t_{\rm end})}{3} S_\sigma \, ,
\end{equation}
where we have neglected terms proportional to $\Omega_\sigma/\Omega_\phi$ and $\Omega_r/\Omega_\phi$. That is, $\zeta$ is copied to $\zeta_\phi$.

The flaton eventually decays into the Standard Model sector. Here we do not consider the curvaton decay contribution to the Standard Model particles as we usually do in the conventional scenario, since the curvaton is assumed to be completely decoupled and serves as CDM. We also assume that the curvaton is not produced directly from the flaton decay.  Expressing the final radiation component by subscript $R$, which is the sum of the initial radiation and the decay products of the flaton, one obtains on the decay hypersurface,
\begin{equation}
\label{eq:beforedecay}
\Omega_\phi e^{3(\zeta_\phi-\zeta)} + \Omega_\sigma e^{3(\zeta_\sigma-\zeta)} + \Omega_r e^{4(\zeta_r-\zeta)} = 1 \, ,
\end{equation}
just before the flaton decay and 
\begin{equation}
\label{eq:afterdecay}
\Omega_\sigma e^{3(\zeta_\sigma-\zeta)} + \Omega_R e^{4(\zeta_R-\zeta)} = 1 \, ,
\end{equation}
just after the flaton decay. Expanding each exponential up to linear order and matching $\zeta$ at the moment of the flaton decay, we obtain
\begin{equation}
\label{zeta_R}
\zeta_R \approx \zeta_\phi \approx \zeta_r + \frac{r_\sigma(t_{\rm end})}{3} S_\sigma \, , 
\end{equation}
where we have used eq.~\eqref{eq:zeta_flaton}. 
Thus, further assuming that the curvature perturbation associated with the original radiation component is negligible as in the usual curvaton scenario, our scenario predicts the positively-correlated CDM isocurvature perturbation as
\begin{equation}
S_{\rm CDM} \equiv 3(\zeta_\sigma-\zeta_R) \approx \frac{3(1-r_\sigma (t_{\text{end}}))}{r_\sigma (t_{\text{end}})} \zeta_R \, .
\end{equation}
As we have seen in eq.~\eqref{eq:r_sigma}, $r_\sigma(t_{\rm end})$ only cares the ratio of $\Omega_r$ and $\Omega_\sigma$ and is independent of the flaton contribution. Thus we can readily realize $r_\sigma(t_{\rm end}) \approx 1$ so that the CDM isocurvature perturbation can be well suppressed while the curvaton itself or its decay product can be a dominant component of CDM.

\begin{figure}[tp]
\centering
\subfigure[$N_{\rm TI} = 7$]{
\includegraphics [width = 0.46\textwidth, clip]{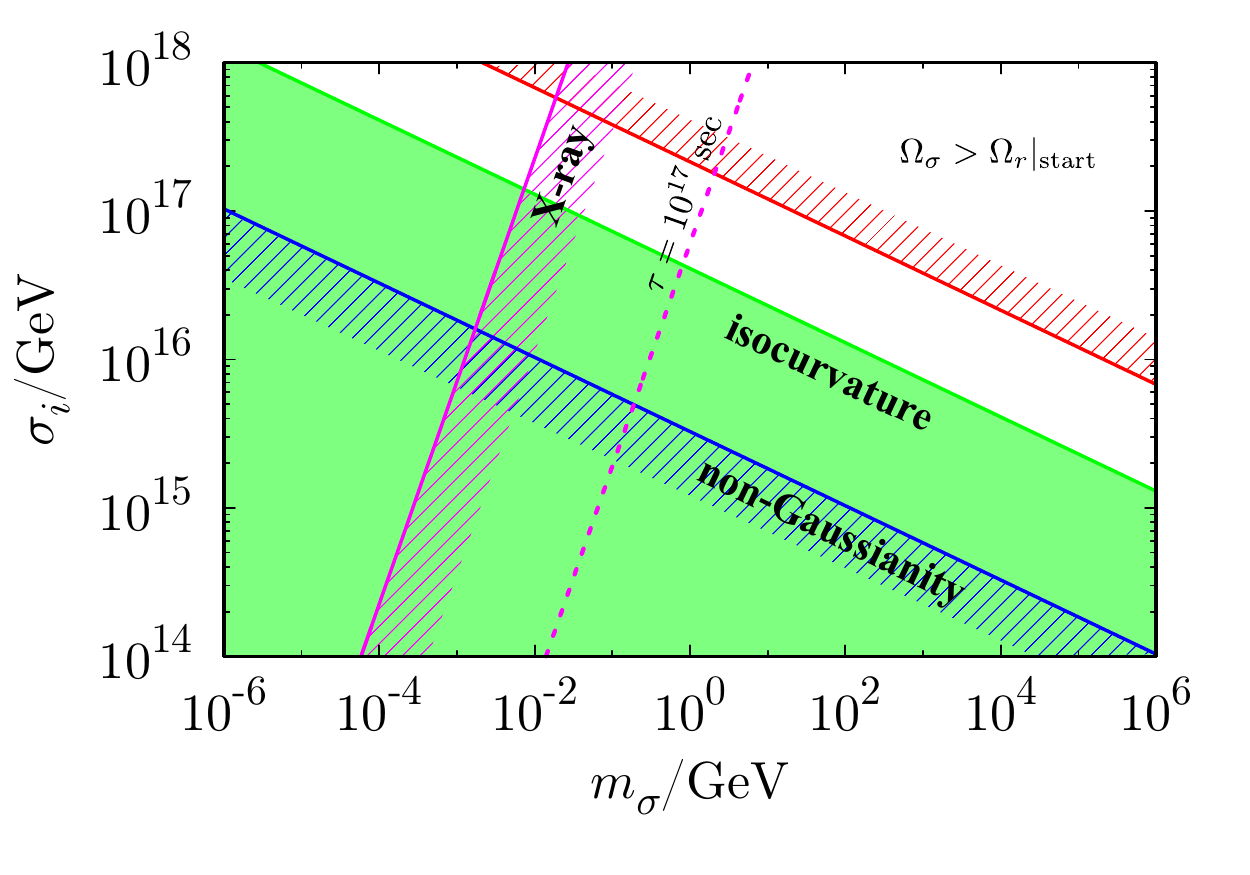}
\label{subfig:case1_a}
}
\subfigure[$N_{\rm TI} = 10$]{
\includegraphics [width = 0.46\textwidth, clip]{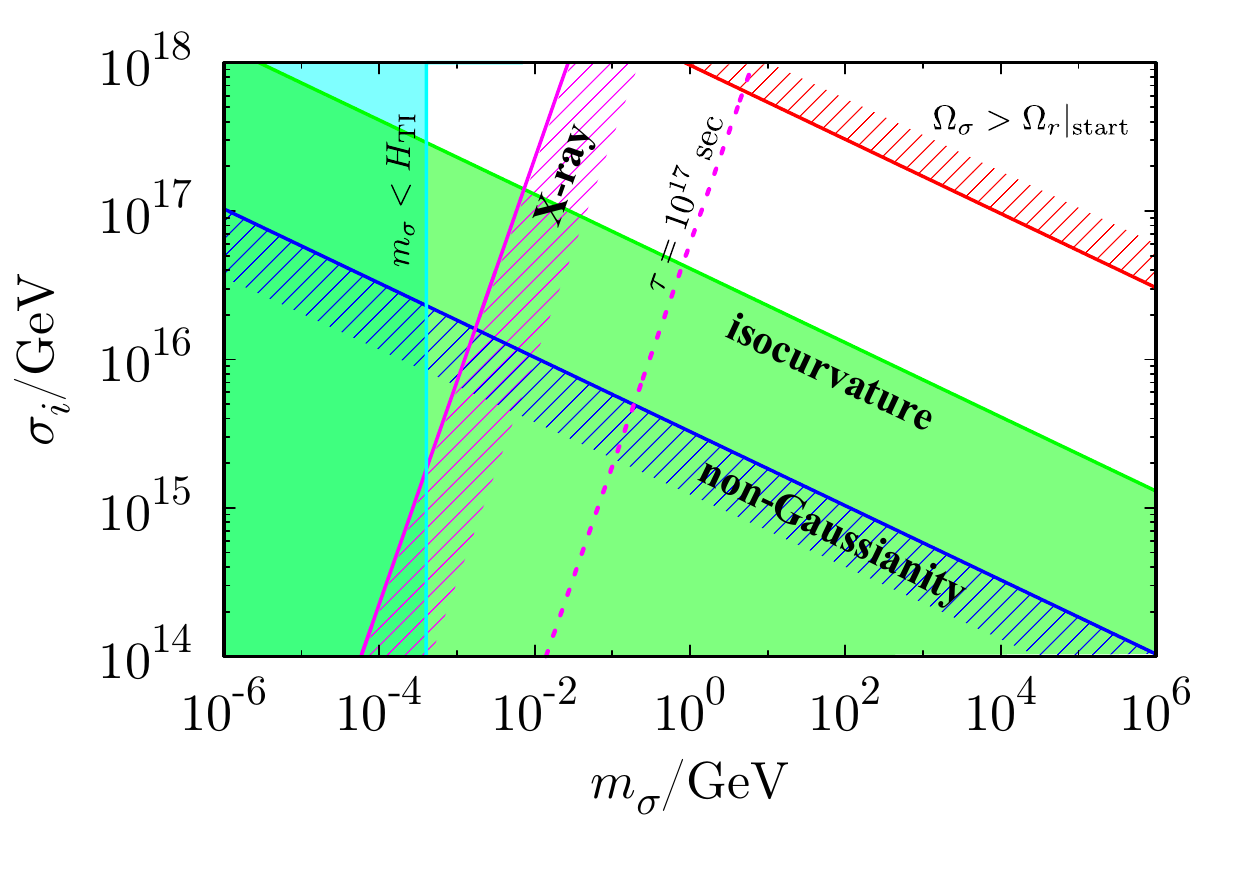}
\label{subfig:case1_b}
}
\caption{
Allowed parameter region for $\Omega_r(t_{\rm start}) > \Omega_\sigma(t_{\rm start})$  [see eq.~(\ref{eq:cdm_abundance_2})] with $N_{\rm TI}=7$ (left panel) and $N_{\rm TI} = 10$ (right panel). We have taken $m_\phi = 1$ TeV. The unshaded regions correspond to the allowed regions in which the curvaton can be the main component of present CDM. The red regions correspond to the other case [$\Omega_\sigma(t_{\rm start}) > \Omega_r(t_{\rm start})$].  This region may or may not be excluded depending on $T_{\text{R}}$ and $N_{\text{TI}}$ (see Figure~\ref{fig:case2}). The green and blue shaded regions express the constraint from the CDM isocurvature and $f_{\rm NL}$ respectively and the cyan region corresponds to $m_\sigma < H_{\rm TI}$, in which the thermal inflation starts before the commencement of the curvaton oscillation. If the curvaton is an axion-like particle, the region below the magenta hatching is also excluded by the constraint from X-ray observation. In this case, the initial amplitude of the curvaton is replaced with the axion decay constant, $\sigma_i = f \theta_i$, with an initial misalignment angle expected to be order one. The  dotted magenta line corresponds to the axion lifetime equal to the present cosmic age.
}
\label{fig:case1}
\end{figure}

\begin{figure}[tp]
\centering
\includegraphics [width = 0.53\textwidth, clip]{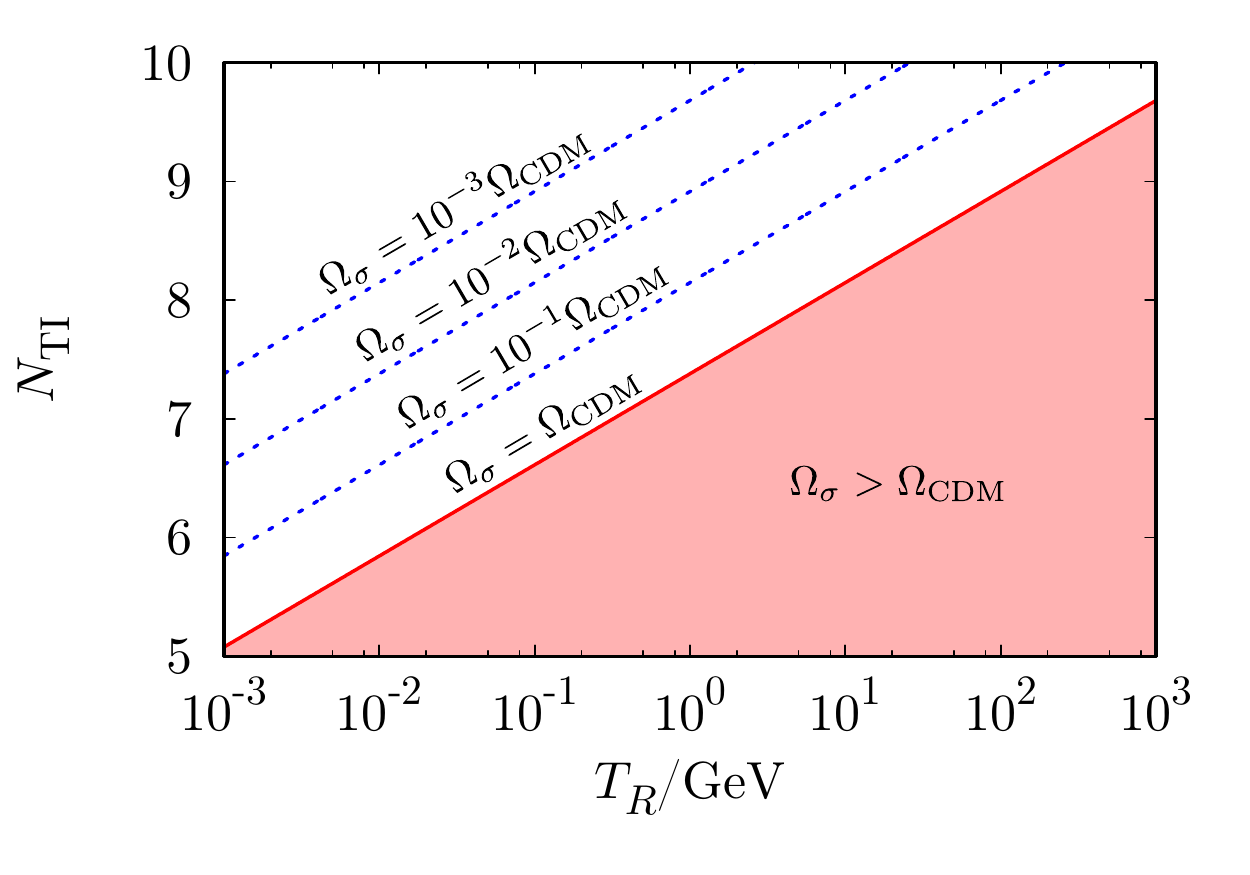}
\caption{
Allowed parameter region for $\Omega_\sigma(t_{\rm start}) > \Omega_r(t_{\rm start})$. [see eq. \eqref{eq:cdm_abundance_2}]. The red region represents the overproduction of CDM. Each blue dotted line shows the contour of equal $\Omega_\sigma$ at present.
}
\label{fig:case2}
\end{figure}

Let us move on to non-Gaussianity. Assuming a quadratic potential for the curvaton $\sigma$, we can express $\zeta_{\sigma}$ in terms of the Gaussian perturbation $\zeta_{\sigma\text{G}} \equiv 2 \delta \sigma / (3\bar{\sigma})$ with $\delta \sigma$ being the fluctuation on an initial flat hypersurface around the spatial average $\bar{\sigma}$ as
\begin{align}
\label{field_perturbation}
\zeta_\sigma = \zeta_{\sigma\text{G}} - \frac{3}{4}\zeta_{\sigma\text{G}}^2 + \frac{3}{4}\zeta_{\sigma\text{G}}^3 + \cdots \, ,
\end{align}
where we have used eq.~\eqref{zeta_X}. The non-linear parameters $f_{\text{NL}}$ and $g_{\rm NL}$ are defined as
\begin{align}
\label{fNL_def}
\zeta = \zeta_{\text{G}} + \frac{3}{5} f_{\text{NL}} \zeta_{\text{G}}^2 
+ \frac{9}{25} g_{\rm NL} \zeta_{\text{G}}^3 + \cdots \, ,
\end{align}
where $\zeta_{\text{G}}$ is the Gaussian part of $\zeta$. Expanding eqs.~\eqref{eq:beforeTI}, \eqref{eq:afterTI} and \eqref{eq:afterdecay} up to third order, we find
\begin{equation}
\label{eq:totalzeta}
\zeta \approx \left[ r_\sigma + \mathcal{O}(\Omega_{\sigma} ) \right] \zeta_{\sigma\text{G}} 
+ \left[ \frac{3}{4}r_\sigma +\mathcal{O}(\Omega_\sigma) \right] \zeta_{\sigma\text{G}}^2
+ \mathcal{O}(\Omega_\sigma) \zeta_{\sigma\text{G}}^3 \, ,
\end{equation}
with the right-hand side being evaluated at the end of thermal inflation. Given that $\Omega_\sigma(t_{\rm end}) \ll 1$, comparing the total curvature perturbation with eq.~\eqref{fNL_def} gives
\begin{align}
f_{\rm NL} = \frac{5}{4r_\sigma(t_{\rm end})} 
\quad \text{and} \quad
g_{\rm NL} = \mathcal{O}(\Omega_\sigma) \, .
\end{align}
The vanishing $g_{\rm NL}$ is distinctively different from the standard curvaton scenario~\cite{Sasaki:2006kq}. Figures~\ref{fig:case1} and \ref{fig:case2} show allowed parameter regions associated with two cases [see eq.~(\ref{eq:cdm_abundance_2})]. Note that the region above the green line in Figure~\ref{fig:case1} is free from both the isocurvature and $f_{\rm NL}$ constraint and realizes the curvaton DM as a main component of the present CDM. In the red region in Figure~\ref{fig:case1}, the opposite case $\Omega_\sigma(t_{\rm start}) > \Omega_r(t_{\rm start})$ is realized, whose exclusion depends only on $T_R$ and $N_{\text{TI}}$.  This is shown in Figure~\ref{fig:case2}.

Before closing the section, we make comments on the secondary inflation. An important feature for this period is that the secondary inflaton is homogeneous and that the associated perturbation is copied from the other components at the end of inflation. The same mechanism to copy $\zeta$ to $\zeta_{\phi}$ may arise for other types of the second (short-term) inflation. For example, let us consider a short-term hilltop inflation. In this case, if the inflaton is initially stabilized e.g.~by a Hubble-induced mass at the origin, the super-horizon perturbation of the inflaton could be generated by the same mechanism. This mechanism works when the super-horizon perturbation of the secondary inflaton is heavily suppressed. In other cases, the inflaton may have its original perturbation $\zeta_{\phi}$, a priori unrelated to $\zeta_{\sigma}$ or $\zeta_{r}$ and it can be responsible for the generation of the curvature perturbation as in the case of the inflating curvaton~\cite{Dimopoulos:2011gb}. But in this case, $\sigma$ has uncorrelated isocurvature perturbations, which may conflict with high scale inflation scenarios. We do not discuss this case in detail. We are mainly interested in a situation like thermal inflation where the eq.~\eqref{eq:zeta_flaton} holds.

\section{String axion with thermal inflation} 
\label{sec:ALP}

In the previous section, we have considered the general possibility that the curvaton $\sigma$ has a long life time and plays the role of CDM. Thus it must have weak enough interactions with the Standard Model particles, and the possible candidates include dilaton or moduli in string theory, and axion-like particles (ALPs) (more generally pseudo Nambu-Goldstone bosons). Among them, string axions or ALPs may be plausible candidate because of the flatness of their potential due to the original shift symmetries at high energy scale. For general hidden sector ALPs, their mass may be dependent on the temperature of the hidden sector thermal bath, if any, but we neglect such model-dependent temperature dependence for simplicity. For an ALP $\sigma$ to have its quantum fluctuation, the corresponding symmetry must be broken during or before the primordial inflation. In addition, in order to be consistent with the CMB data, $f \gtrsim H_{\rm inf}$ is necessary, where $f$ is the decay constant of the ALP. For definiteness, let us consider a string axion. The potential is given by
\begin{equation}
\label{eq:cos-V}
V(\sigma) = \Lambda^4 \left[ 1 - \cos \left( \frac{\sigma}{f} \right) \right] \, ,
\end{equation}
where $\Lambda$ is some energy scale associated with the shift symmetry breaking. The axion mass is given by $m_\sigma = \Lambda^2/f$ and the final abundance is given by eq.~\eqref{eq:cdm_abundance} with replacing $\sigma_i = f \theta_i$, where $\theta_i$ is the initial misalignment angle.

Axions can interact with the Standard Model vector bosons due to the corresponding anomalies depending on the assignment of quantum numbers in the model. In particular, it can couple to photon generically through e.g. heavy quark loops, and it may produce observable signals. Such a coupling is described by the interaction term
\begin{equation}
{\cal L}_{\sigma \gamma \gamma} = \frac{\alpha_{\rm EM}}{4 \pi} \frac{\sigma}{f} F_{\mu\nu} \tilde{F}^{\mu\nu} = \frac{g_{\sigma\gamma\gamma}}{4} \sigma F_{\mu\nu} \tilde{F}^{\mu\nu} \, ,
\end{equation}
where $F_{\mu\nu}$ is the field strength tensor of photon, $\tilde{F}^{\mu\nu} = \epsilon^{\mu\nu\rho\sigma} F_{\rho\sigma}/2$ is its dual and $\alpha_{\rm EM}$ is the fine structure constant of electromagnetic interaction. In the above formula, a possible model-dependent coefficient of ${\cal O}(1)$ has been absorbed by the redefinition of $f$ or $g_{\sigma\gamma\gamma}$. The decay rate of the axion into two photons is
\begin{equation}
\Gamma_{\sigma \to \gamma\gamma} = \frac{g^2_{\sigma\gamma\gamma}}{64 \pi} m_\sigma^3 \, .
\end{equation}
If the lifetime $\tau = 1/\Gamma_{\sigma \to \gamma\gamma}$ is longer than the age of the universe, the axion can contribute to the present CDM. The lifetime is estimated as
\begin{equation}
\label{eq:lifetime}
\tau \approx 3 \times 10^{24}~{\rm sec} \times 
 \left( \frac{m_\sigma}{1~{\rm MeV}} \right)^{-3} \left( \frac{f}{10^{16}~{\rm GeV}} \right)^2 \, .
\end{equation}
It puts an upper bound on the axion mass. More stringent constraints come from diffuse X($\gamma$)-ray observations \cite{Essig:2013goa}. In Figure~\ref{fig:case1}, the regions right to the magenta solid line (from X($\gamma$)-ray) and the magenta dotted line (from lifetime) are excluded for the ALP curvaton-CDM. Our scenario favors a large decay constant $\gtrsim 10^{17}$ GeV which would result in the overproduction and/or the isocurvature problem in the standard case. Note that the axion in the allowed region but close to the X($\gamma$)-ray contour can be a decaying dark matter and predict an X-ray line excess~\cite{Higaki:2014zua}.

\section{Conclusions and discussions}
\label{sec:conclusion}

In this article, we have considered a simple and economical scenario in which a long-lived curvaton plays the role of CDM. A necessary ingredient is a short, secondary inflation that dilutes curvaton energy density as well as other relic density and produces radiation-dominated universe after the second reheating. Furthermore, demanding that the secondary inflaton has no intrinsic initial fluctuations and the transition between the inflationary and the oscillation phases occurs suddenly, the perturbation in the curvaton sector is copied to that in the secondary inflaton. In such a way, both the overproduction of unwanted relics and a large CDM isocurvature perturbation can be evaded. Although we have considered thermal inflation explicitly as the secondary inflation, any type of inflation satisfying the above conditions can work for the curvaton-dark matter scenario.

As a concrete realization, we have considered a string axion as the curvaton-dark matter candidate. Taking into account the observational bounds on the CDM abundance, CDM isocurvature perturbation, non-Gaussianity and diffuse X-ray signal, we have confirmed that there is a viable window for the axion parameters satisfying $m_\sigma \lesssim {\cal O}(10)$ MeV and $f \sim 10^{17}$ GeV.

Finally, we comment on the possibility of identifying the QCD axion as our curvaton-dark matter. Naively the QCD axion is not compatible with our scenario, because its mass is too tiny.  Since the axion needs to start  oscillations before or during the secondary inflation for our scenario to work well, the flaton mass must then also be a tiny value $\lesssim 1$ GeV.  This is quite unrealistic especially in the models motivated by supersymmetry (SUSY).  This issue can be addressed by various scenarios which make the axion heavy.  For example, the axion mass can obtain an additional contribution in the presence of an extra fermion of a high dimensional color representation~\cite{Kobakhidze:2016wmv} or a number of extra vector-like quarks~\cite{Kobakhidze:2016rwh} (see also references therein for related early attempts).  The additional mass contribution can come from chiral symmetry breaking in a hidden sector without spoiling the solution to the strong CP problem utilizing extension of the color group~\cite{HA_colorgroup} or $\mathbb{Z}_2$ exchanging parity with the mirror sector~\cite{HA_Z2} (see also Refs.~\cite{cp-sol}).

Alternatively, the axion mass at the vacuum may be completely the standard one but it might have been heavier in the early universe~\cite{HA_temporary, Choi:2015zra}.  The origin of such a time-dependent additional mass is again model-dependent, but let us focus on the (SUSY) ``stronger QCD'' scenario in Ref.~\cite{Choi:2015zra} as an example, in which the chiral symmetry breaking scale (and hence the axion mass) is field/Hubble-dependent.  During the primordial epoch with sufficiently large $H$, the electroweak symmetry is supposed to be broken badly, which makes quarks heavy.  This in turn changes the renormalization group running of the strong coupling, and the QCD scale becomes temporarily high.  The axion mass can be raised to $m_a \sim 0.1-1$ MeV for $m_{\text{SUSY}} < T < \widetilde{\Lambda}_{\rm QCD}$ where $m_{\rm SUSY} \sim {\cal O}(1)$ TeV and $\widetilde{\Lambda}_{\rm QCD} \sim \mathcal{O}(10)$ TeV is the temporarily enhanced QCD scale~\cite{Choi:2015zra}. Interestingly, the secondary inflation naturally occurs before the process that Higgs transits to the true electroweak vacuum.  On the other hand, the axion gets an additional misalignment at this transition, and an additional care is required to estimate its abundance. Combined with our scenario, this scenario may provide us the QCD axion as the curvaton-dark matter.  We will leave it as an interesting future study.

\subsection*{Acknowledgment}

We acknowledge the support from the Korea Ministry of Education, Science and Technology, Gyeongsangbuk-Do and Pohang City for Independent Junior Research Groups at the Asia Pacific Center for Theoretical Physics.
JG is also supported in part by a TJ Park Science Fellowship of POSCO TJ Park Foundation, and by the Basic Science Research Program through the National Research Foundation of Korea (NRF) Research Grant NRF-2016R1D1A1B03930408. 
TT is partially supported by the Grant-in-Aid for JSPS Fellows and the Grant-in-Aid for Scientific Research on Scientific Research No.~26$\cdot$10619, and by National Research Foundation of Korea (NRF) Research Grant NRF-2015R1A2A1A05001869.  A part of TT's work was done when he belonged to APCTP.

\end{document}